\documentclass[dvips]{acta}
\usepackage{supertabular,lscape,epsfig}
\usepackage{amssymb}
\usepackage{amsmath}
\usepackage{txfonts}

\SetPages{0}{0}

\SetVol{00}{2014}

\begin{document}

\begin{Titlepage}

\Title{The study of triple systems V819~Her, V2388~Oph, and V1031~Ori} \Author{Zasche, P.$^{1}$,
Uhl\'a\v{r}, R.$^{2}$, Svoboda, P.$^{3}$}
 {$^{1}$ Astronomical Institute, Charles University Prague, Faculty of Mathematics and Physics,
 V Hole\v{s}ovi\v{c}k\'ach 2, Praha 8, CZ-180 00, Czech Republic\\
e-mail: zasche@sirrah.troja.mff.cuni.cz \\
 $^{2}$ Private Observatory, Poho\v{r}\'{\i} 71, J\'{\i}lov\'e u Prahy, CZ-254 01, Czech Republic\\
 $^{3}$ Private observatory, V\'ypustky 5, Brno, CZ-614 00 Czech Republic}

\Received{Month Day, Year}
\end{Titlepage}

\Abstract{The systems V819~Her, V2388~Oph, and V1031~Ori are triples comprised of an eclipsing
  binary orbiting with a distant visual component on a longer orbit. A detailed analysis of these
  interesting systems, combining the two observational techniques: interferometry and apparent period variation,
  was performed.
The interferometric data for these three systems obtained during the last century determine the
  visual orbits of the distant components in the systems. The combined analysis of the positional
  measurements together with the analysis of apparent period changes of the eclipsing binary (using the minima
  timings) can be used to study these systems in a combined approach, resulting in a set of parameters
  otherwise unobtainable without the radial velocities.
The main advantage of the technique presented here is the fact that one needs no
  spectroscopic monitoring of the visual orbits, which have rather long periods: 5.5 yr for V819~Her, 9.0~yr
  for V2388~Oph, and 31.3~yr for V1031~Ori, respectively. The eccentricities of the outer orbits
  are 0.69, 0.33, and 0.92, respectively. Moreover, the use of minima timings of the eclipsing pairs
  help us to derive the orientation of the orbit in space with no ambiguity around the
  celestial plane. And finally, the combined analysis yielded also an independent determination of
  the distance of V819~Her ($68.7 \pm 1.8$~pc), and V2388~Oph ($70.6 \pm 8.9$~pc).
We also present a list of similar systems, which would be suitable for a combined analysis like
  this one.}
{binaries: eclipsing -- binaries: visual -- stars: fundamental parameters -- stars: individual:
V819~Her, V2388~Oph, V1031~Ori.}

\section{Introduction}

Multiple stellar systems (i.e. of multiplicity three and higher) are excellent objects to be
studied. The importance of such systems lies in the fact that we can study the stellar evolution in
them, their origin, tidal interaction, testing the influence of the distant components to the close
pair, Kozai cycles, studying the dynamical effects and precession of the orbits, but also the
statistics and relative frequency of such systems among the stars in our Galaxy (and outside), see
e.g. Tokovinin (2007), Guinan \& Engle (2006), or Goodwin \& Kroupa (2005).

A few years ago we introduced (Zasche \& Wolf 2007) a new method of combining
the different observational techniques
:  namely the analysis of the visual orbit (positional measurements of double stars obtained via
interferometry, in former times via micrometry) together with the apparent period changes of the
eclipsing pair (study of times-of-minima variation), into one simultaneous fitting procedure. This
approach deals with very favourable triple systems, where one of its components is an eclipsing
binary, while the distant component is being observed via interferometry. In general, for the
triple systems the orbital periods of the wide orbits are usually the crucial issue. The shorter
periods are mostly being discovered via spectroscopic monitoring, while the longer ones via
interferometric detection of the distant components. The space in between these two methods is
often harvested via a so-called Light-Time Effect (hereafter LITE, see e.g. Irwin 1959), which is
able to discover the orbits of components with orbital periods from hundreds of days to hundreds of
years (Zakirov 2010). The whole method is using a well-known effect of detecting apparent periodic
shifting of the eclipsing binary period as the eclipsing pair revolves around a barycenter with the
third component. In principle, every triple system with an eclipsing binary should display some
amount of apparent periodic modulation of the inner eclipsing period (the only exception is the
face-on orientation of the wide orbit). Hence, a long-term monitoring of such systems can be very
fruitful. Therefore, in 2006 we started photometric monitoring of selected systems, resulting
partly in the present paper.

Raghavan et al. (2010) published their results about the multiplicity fraction of the solar-type
stars as compared with previous results on different spectral types (showing that the multiplicity
fraction rapidly decreases to lower mass stars). Hence, one can expect that also the number of
multiples within the group of eclipsing binaries would be large. On the other hand, there is still
a limited number of such systems, for which both inner and outer orbits are known. The problem is
usually the period of the outer orbit (sometimes of the order of hundreds of years), and
surprisingly also the brightness of the systems. The stars observed via interferometry need to be
rather bright (usually $<$10mag), but for such bright targets the photometry is often hard to
obtain because these can easily saturate the modern CCD detectors mount on even modest telescopes.
And finally, also the presence of the third component makes the eclipses of the inner pair more
shallow due to its brightness (the third component often cannot be separated and is also observed
in the aperture). All of these reasons make such triple systems rather rare and unusual, and
currently we still know only about 30 visual multiple systems with eclipsing components for which
both orbits are known (see e.g. Zasche et al. 2009).

\section{The approach for the analysis}

From the potentially interesting systems suitable for the combined analysis, we have chosen three
stars, following the criteria introduced here. All of the triple stars are rather bright: $V$ =
5.57~mag for V819~Her, $V$ = 6.26~mag for V2388~Oph, and $V$ = 6.06~mag for V1031~Ori; all of them
are observable from our observatories in the Czech Republic, all have rather deep minima (which
make them suitable for small telescopes), and all have rather short periods of the visual
orbits (5.5 to 31.3 years).

For the three selected stars, we used the following approach. At first, all of the available
interferometric observations were collected and analysed. These measurements are stored e.g. in the
``Washington Double Star Catalog"\footnote{http://ad.usno.navy.mil/wds/} (hereafter WDS, Mason
et~al. 2001), together with the orbital information (if available). Analysing the complete set of
observations, we obtained an updated solution of the visual orbit and the set of its parameters $(
a, p, i, e, \omega, \Omega, T )$, where $a$ is the semimajor axis, $p$ is the period of the outer
orbit, $i$ is its inclination, $e$ the eccentricity, $\omega$ the argument of periastron, $\Omega$
the longitude of the ascending node, and $T$ the time of periastron passage. The least-squares
method and the simplex algorithm was used (see e.g. Kallrath \& Linnell 1987) for computing.

With this solution, we collected all available times of minima observations, mostly stored in the
online database\footnote{http://var.astro.cz/ocgate/}, Paschke \& Br\'at (2006). Many new
observations for these three systems were also obtained and the times of minima derived by the
authors, see the tables in the Appendix section. Having a preliminary solution of the visual orbit
and plotting the available times of minima in the $O-C$ diagram, one can easily judge whether a
system is suitable for a combined analysis or not.

The selection mechanism was rather easy - sufficient coverage of the long orbit at least in part of
its period, in the best case the whole period $p$ covered by both methods, especially during the
periastron passages. With the values of parameters derived from the visual orbit, the times
of minima were analysed using the LITE hypothesis (see e.g. Mayer 1990). 
This led to the set of the LITE parameters $(p, A, e, T, \omega_{1}, JD_0, P)$, where $JD_0$ and
$P$ are the linear ephemerides of the eclipsing pair, and $A$ is the amplitude of the LITE. This
amplitude tells us what is the magnitude of the delay caused by the third body, hence its value is
largely affected by the mass of the third body and inclination between the orbits. The angles
$\omega_1$ derived from the LITE and $\omega$ derived from the visual orbit can be the same, or can
be shifted by $180^{\circ}$. This ambiguity in argument of periastron is due to the fact that we
usually do not know which of the two components on the visual orbit is the eclipsing binary
(because we cannot separate the two stars photometrically). However, the combined solution is able
to solve this and only one $\omega$ value is computed in the code. The other possibility is to
measure the radial velocities on the long orbit.

If the system was found to be suitable for a combined approach, we used the code introduced in our
previous work (Zasche \& Wolf 2007). The starting values of the fit are the preliminary values of
parameters as derived for separate solutions. After several iteration steps using the simplex
algorithm, the final and acceptable solution was reached. This solution usually gives a fair and
acceptable solution for both our data sets for a particular system. The whole computational
procedure is stable and converge rather rapidly when both methods have good orbital coverage. If
this is not the case and the fitting process provides spurious results, we can also proceed
iteratively and fit only some parameters, not to let the programme compute all the parameters
simultaneously in one step.

There still remains an open question - of how the amplitude $A$ of the LITE and the semimajor axis
$a$ of the visual orbit are connected? Here comes the most important step in our approach, a way
how to derive the distance to the system. If we know the distance to the triple (which is usually
true, because the stars are bright and close and were mostly observed by the Hipparcos satellite),
both $a$ and $A$ are directly connected. The individual masses of all three components are known,
because we know the eclipsing binary masses from the light curve and radial velocity curve solution
(from the already published papers) and the total mass of the whole system from the visual orbit.
Hence, using our combined analysis yielding also both $A$ and $a$ values, the distance to the
multiple system can be easily derived independently of other techniques. The detailed description
of how the values are connected and the distance can be derived is given in Zasche \& Wolf (2007).
The same approach was used in Zasche \& Wolf (2007) for the system VW~Cep, and also for KR~Com in
Zasche \& Uhl{\'a}{\v r} (2010).

Our final fit for all systems was obtained minimizing the total chi-square value, which is being
easily computed as a sum of both chi-square values of LITE together with the chi-square value of
the visual orbit (see e.g. Zasche (2008) for discussion).

\subsection{The weightening scheme}

The problem with the individual uncertainties in both methods was solved in the following way. The
individual errors of the minima times observations were used (when available) for computing, only
in these cases where this information was missing we assumed an artificially high value of 0.01~day
uncertainty. This was usually the case of old photoelectric data, where the error was not published
in the original paper.

Because of unavailable error estimation for the older astrometric data, we have to use a different
approach. For the positional measurements the individual weights were used instead of the
uncertainties. These very different techniques provide us with an order of magnitude different
precision when deriving the positions of the two components, despite the fact that the technique is
called in general ``interferometric'' (e.g. visual interferometry versus long-baseline Palomar
Testbed Interferometer). Therefore, the weightening scheme was the following:
  \begin{itemize}
    \item 1 -- Visual interferometer\\[-7mm]
    \item 5 -- Interferometric technique (phase grating interferometer)\\[-7mm]
    \item 5 -- Hipparcos satellite\\[-7mm]
    \item 10 -- Speckle interferometric technique (CHARA speckle, USNO speckle)\\[-7mm]
    \item 10 -- Adaptive optics\\[-7mm]
    \item 100 -- Long-baseline visual/IR/radio interferometer (Palomar Testbed Interferometer)\\[-7mm]
 \end{itemize}


As is stated below, the individual published errors of the times of minima are usually
underestimated. Artificially increasing these values we tested how the final fit changes.
Obviously, the larger the errors in one method (times of minima analysis) -- the closer the final
fit to the other method (visual orbit). This was tested, but the finding was that doubling the
errors of the data points in the minima times data set led to only slight change of the final fit
and its parameters (not more than 2-3\,\%).

\section{Individual systems}

\subsection{V819 Her}

V819 Her (= HD~157482 = HR~6469 = HIP~84949 = MCA~47) is an Algol-type eclipsing binary with its
magnitude about 5.57 in V filter. The photometric analysis was performed by van Hamme et al.
(1994), who analysed the color indices and derived the individual spectral types as F2V and F8 for
the eclipsing pair, while G8 IV-III for the third component, similar result was also found by
Scarfe et al. (1994) analysing the spectra of the system. The third component is also
photometrically variable, probably due to the spots on its surface. The eclipsing pair is orbiting
around a common center of mass with a third star in a 5.5-yr period visual orbit, having the
eccentricity of about 0.67, and the LITE due to this movement is evident. The wider pair was
discovered by speckle interferometry in 1980 (McAlister et~al. 1983) and has been extensively
observed by this technique and more recently with the Palomar Testbed Interferometer (Muterspaugh
et~al. 2008), and also using the CHARA Array (O'Brien et al. 2011). In addition, the eclipsing pair
was resolved (Muterspaugh et~al. 2006), yielding also the mutual inclination of both orbits as
33.5$^\circ$ (most recently by O'Brien et al. 2011).

This is the only system in our sample where the LITE was analyzed together with the interferometry
and radial velocity data before (see Muterspaugh et~al. 2006). The first analysis of the LITE
together with the visual orbit was that by Wasson et al. (1994), who also observed many times of
minima over a 10-year period. However, since then a lot of new measurements were obtained, so a new
up-to-date analysis would be very profitable, especially when dealing with our new unpublished
photometric observations.

The measurements of the eclipses of the inner pair were carried out during the epoch 2008--2014 by
two of the authors (RU and PS) at their private observatories in the Czech Republic and northern
Italy. These data were mostly obtained by small 35-mm cameras equipped with CCD detectors. Such
small instruments were used because of the high brightness of the target. All of the observations
were routinely reduced with dark frames and flat fields. The resulting times of minima were
calculated using a standard Kwee-van Woerden method (Kwee \& van Woerden 1956). The new times of
minima are given in the Appendix tables together with the already published ones. The errors of the
individual observations are given together with the type, filter and reference.

\begin{figure}
  \centering
  \includegraphics[width=\textwidth]{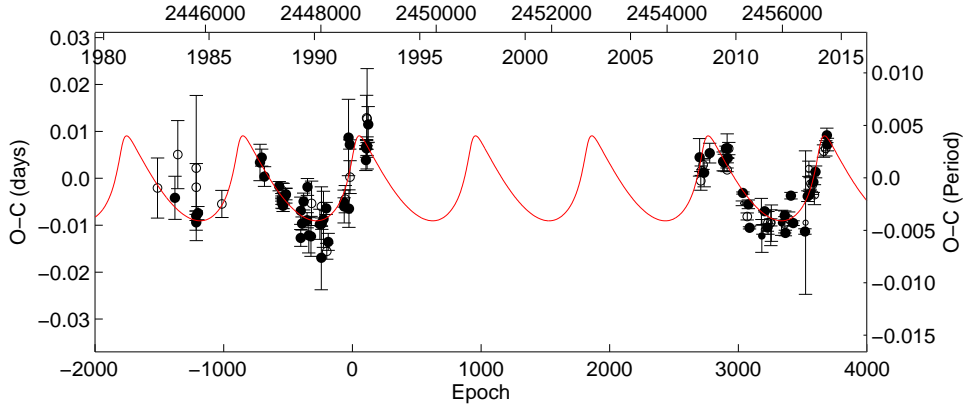}
  \caption{$O-C$ diagram of V819 Her, plotted with all available times of minima observations. The dots stand
  for the primary, while the open circles for the secondary minima. The solid curve represents the final LITE
  fit as derived from the combined solution of the triple. The larger symbols stand for the observations
  with the higher precision.}
  \label{FigV819HerOC}
\end{figure}

\begin{figure}
  \centering
  \includegraphics[width=100mm]{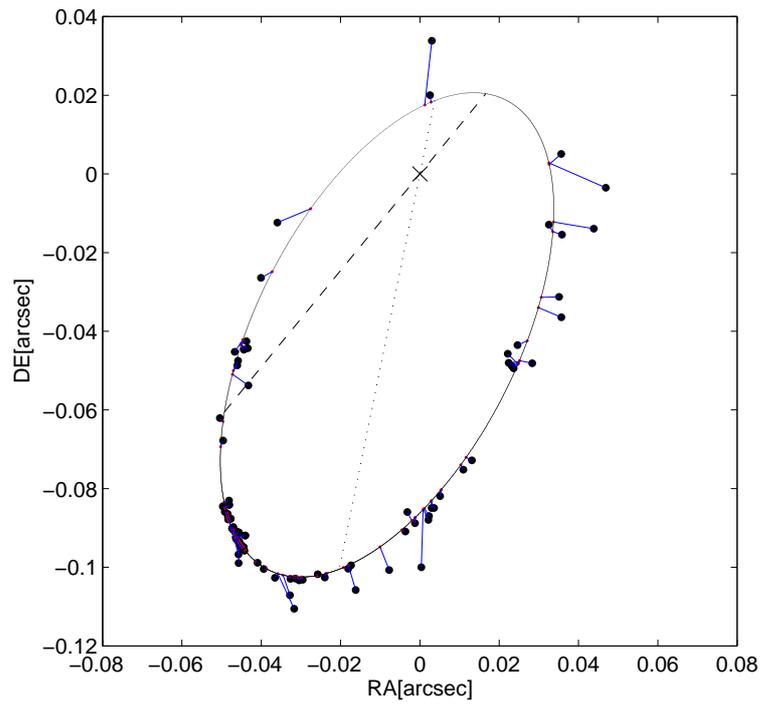}
  \caption{Visual orbit of V819 Her as projected on the sky. The individual observations are plotted as dots,
  connected with their theoretical positions on the orbit with short abscissae. The dashed line represents
  the line of the apsides, while the dotted one stands for the line of the nodes. The eclipsing
  binary is located in the coordinates (0,0).}
  \label{FigV819HerOrbit}
\end{figure}

We used the same computational approach as introduced in Section 2. In total, 102 times of minima
(of which 43 are our new unpublished data) were used for the analysis, together with 114
interferometric measurements of the visual double. The observations of minima times obtained in
different filters during one night, were averaged into one point. All the data points used are
given in the Appendix section Tables.

The analysis started with the values of parameters as presented in Muterspaugh et~al. (2008) and
then followed the procedure as described above. The resulting fits of the combined analysis are
plotted in Figs. 1 and 2 and the parameters are given in Table 1. As one can see, the final fit of
the 5.5-yr variation is clearly visible in Fig 1, in agreement with the visual orbit plotted in
Fig. 2. The 5.5-yr period is now well-defined, and also the periastron passages are well-covered,
hence we can also try to calculate the distance to the system.

In Fig. 1 there is slightly larger scatter of the data points than one would expect from the
individual error bars plotted. It is probably caused by a photometric variability of the third
component, which could also influence the shape of both primary and secondary minima. Hence, also
the times of minima and precision of their derivation can be affected. The magnitude of $O-C$
scatter caused by a presence of spots was studied e.g. by Watson \& Dhillon (2004), who found it to
be well below our minima precision, of the order of seconds only. On the other hand, our recent
study of two binaries with asymmetric light curves and application of standard Kwee-van Woerden
method for minima times derivation led to rather different finding that this can be up to 15
minutes (Zasche 2011) for very asymmetric light curves. Nevertheless, as one can see from Fig. 1,
the individual error bars are rather underestimated. This is usually caused by a fact that the
errors of minima are usually only the formal errors as derived from the Kwee-van Woerden method
itself, but real uncertainty should be order of magnitude larger. On the other hand, it is
noteworthy that the scatter of our new observations is lower than the scatter of the older data
from 1980's and 1990's, despite the fact that we were using an order of magnitude smaller
telescopes, but equipped with CCD cameras instead of photoelectric photometers.

\begin{table}

 \begin{minipage}{130mm}
 \centering
  \caption{Final parameters of combined solution for V819~Her. The errors are given in the
  parenthesis, as resulted from the program.}  \label{ParametersV819Her}
  \begin{tabular}{@{}c c c@{}}
\hline
                     &   \multicolumn{2}{c}{V819 Her}  \\
 Parameter           &   Present study   & Muterspaugh et~al. (2008) \\
 \hline
 $JD_0 $             & 2448546.5954 (7)  &  2452627.17 (0.29)  \\
 $P$ [d]             & 2.2296330 (19)    &  2.2296330 (19)     \\
 $p$ [day]           & 2015.8 (54.9)     &  2019.66 (0.35)     \\
 $p$ [yr]            & 5.519 (0.150)     &  5.530 (0.001)      \\
 $A$  [day]          & 0.0090 (8)        &  0.0088 (9) $^a$    \\
 $T$                 & 2452621.6 (81.2)  &  2452627.5 (1.3)    \\
 $\omega$ [deg]      & 223.8 (3.5)       &  222.50 (0.22)      \\
 $e$                 & 0.687 (11)        &  0.6797 (7)         \\
 $i$ [deg]           & 56.82 (1.63)      &  56.40 (0.13)       \\
 $\Omega$ [deg]      & 140.8 (5.9)       &  141.96  (0.12)     \\
 $a$ [mas]           & 74.6 (3.7)        &  74.4 (0.9) $^a$    \\ \hline
 $f(m_3)$ [M$_\odot$]& 0.193 (22)        &  0.177 (30) $^a$    \\
 $M_{3}$ [M$_\odot$] & 1.86 (0.30)       &  1.799 (0.098)      \\
 $a_{12}$ [AU]       & 2.16 (0.15)       &  2.11 (0.08) $^a$   \\
 $a_{3}$ [AU]        & 2.97 (0.20)       &  2.99 (0.16) $^a$   \\
 $d$ [pc]            & 68.8 (1.8)        &  68.65 (0.87)       \\
 \hline
\end{tabular}
\end{minipage}
 \centering
  Remark: $^a$ - calculated from the original values
\end{table}

In Table 1 we compare our present results with the previous values of parameters as given in
Muterspaugh et~al. (2008). Some of the parameters were calculated from the published values, just
to be compared with our results. As one can see, the agreement is quite high for most of the
parameters, but the errors given by Muterspaugh et~al. (2008) seem to be rather optimistic. On the
other hand, we present only our formal mathematic errors as resulted from the computational code,
which can be slightly different from a more realistic physical errors. 
However, the superb precision of the results published by Muterspaugh et~al. (2008) is due to the
fact that they analysed the radial velocities and interferometry of a state of the art quality,
covering several periods of the outer orbit. Hence, the Table 1 can serve as a demonstration of our
method based on rather different approach and providing comparable results, with no need of radial
velocities on the long orbit.

As mentioned above, we also tried to compute the distance to the system from our combined ``period
variation -- visual orbit method". Also this value (about 68.8~pc) resulted in very good agreement
with the previously derived values, with no need of long-term spectroscopic monitoring. This value
confirms the previous findings by other authors (e.g. Muterspaugh et~al. 2008, Scarfe et al. 1994,
or O'Brien et al. 2011) that the Hipparcos value $74.0 \pm 4.8$~mas (van Leeuwen 2007) is a bit
shifted.

\subsection{V2388 Oph}

V2388 Oph (= HD~163151 = HIP~87655 = FIN~381) is a W~UMa-type eclipsing binary, a bit fainter than
V819~Her, of about 6.3~mag in $V$ filter. It was discovered as a variable star relatively late, in
1995 by Rodr{\'{\i}}guez et al. (1998). They also classified the star as a W~UMa contact type
(despite incorrectly classified as $\beta$~Lyr on SIMBAD), with the individual temperatures 6450
and 6130~K for both components and an orbital period of the eclipsing pair of about 0.8~days. Both
minima are more than 0.2~mag deep, hence it is an easy target to be observed. Later, Rucinski et
al. (2002) observed the star spectroscopically and classified both components as F-types.

The system was discovered as a visual binary star by Finsen (1963), and since then many new
observations of the pair were carried out. Baize (1988) published its slightly eccentric orbit with
period about 8.3~yr. Later, {S{\"o}derhjelm (1999) gives a better visual-orbit solution with period
8.9~yr. And the most recent study by Docobo \& Andrade (2013) presented the best solution of the
data covering more than 50 years with period 9.008~yr and eccentricity 0.327.

Since its discovery as a variable star, many observations of both primary and secondary minima of
the eclipsing pair were obtained. The light curve analysis has been carried out a few times. The
last one is that by Yakut et al. (2004), who observed the system in $BVR$ filters and analysed the
light curve. Moreover, they also tried to combine the visual orbit parameters as published by
{S{\"o}derhjelm (1999) and the apparent period variation of the eclipsing binary applying the LITE
hypothesis. However, their analysis is incorrect, leading to the large disagreement with the
implied masses from the distance and visual orbit. The problematic point of their analysis was the
fact that they have only poor coverage of the third-body period (only 9 times-of-minima data
points) and they only fit the amplitude of the LITE, fixing all other parameters. Moreover, their
light curve analysis is dubious because of the fact that their fit did not show any total eclipse
(in contradiction with the analysis by Rodr{\'{\i}}guez et al. 1998), but our observations clearly
show that there is a total eclipse lasting more than one hour (see Fig. 3). Hence, their
inclination angle for the eclipsing pair should be higher, closer to the 90$^\circ$, yielding
different masses of the components.

\begin{figure}
  \centering
  \includegraphics[width=85mm]{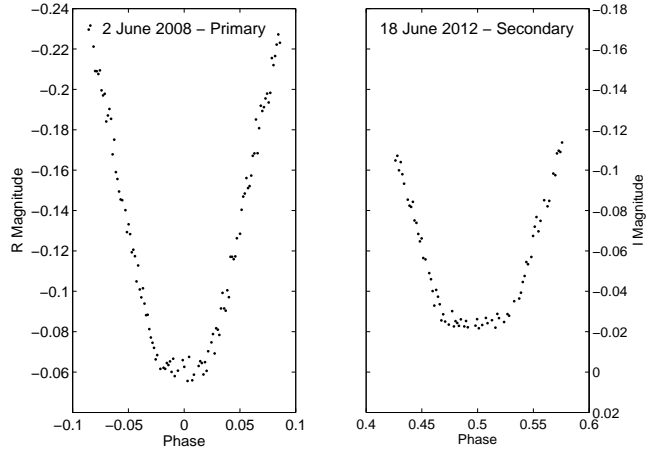}
  \caption{Two observations of minima of V2388 Oph, clearly showing the total eclipses.}
  \label{FigV2388OphLC}
\end{figure}

At this point, our present analysis seems to be much more effective. We evaluated a much larger
data set of minima observations (almost three times longer time base, more than 70 new observations
carried out and reduced by the authors), and our fitting procedure incorporates all relevant
parameters for the combined fit of the visual orbit together with the period variation of the
eclipsing pair. Our analysis is based on 93 minima times observations (see the Appendix tables),
mostly our new ones, which were analysed following the procedure as described in Section 2. The
starting values for the parameters are the ones presented by Docobo \& Andrade (2013). The final
solution of our combined fit is presented in Figs. 4 and 5. The values of all parameters are given
in Table 2. The scatter of the minima times is quite large, but this is probably the physical
scatter of the observations and maximum what can be obtained from the small ground-based
telescopes, because the individual observations were obtained using different instruments,
reduction, weather conditions, etc. To repeat once again, the errors of most of the already
published observations are usually underestimated and only formal ones as derived from the Kwee-van
Woerden method. There is also a possibility of a kind of chromospheric activity (Yakut et al.
2004), which is quite common in this kind of late-type contact binaries. Moreover, the spot on the
surface (Rodr{\'{\i}}guez et al. 1998) makes the whole curve asymmetric, which also brings some
difficulties when deriving the times of minima, see the discussion on the $O-C$ precision in the
previous section.

\begin{figure}
  \centering
  \includegraphics[width=\textwidth]{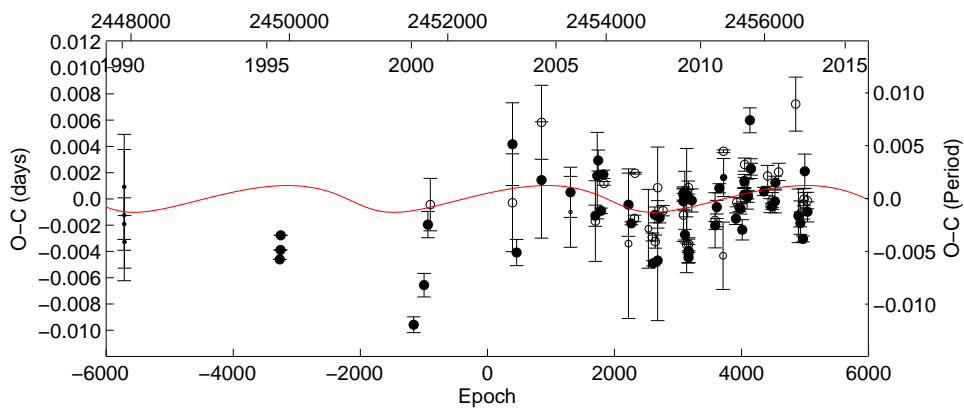}
  \caption{$O-C$ diagram of V2388 Oph. See Fig. 1 for description of the symbols.}
  \label{FigV2388OphOC}
\end{figure}

\begin{figure}
  \centering
  \includegraphics[width=100mm]{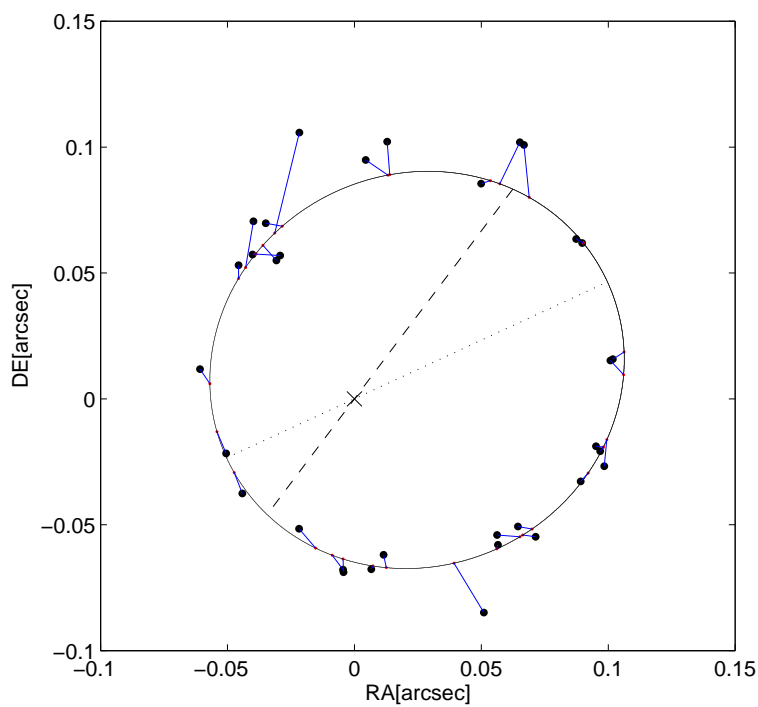}
  \caption{Visual orbit of V2388 Oph.}
  \label{FigV2388OphOrbit}
\end{figure}

Despite large scatter of the minima, we also tried to compute the distance to the system from the
values $a$ and $A$. However, the total mass of the eclipsing pair is necessary for the calculation.
This value was derived from the mass function as derived from the radial velocities by Rucinski et
al. (2002) and using the inclination presented by Rodr{\'{\i}}guez et al. (1998). From this value
of $M_{12} = 1.96 \pm 0.03$~M$_\odot$ we derived the mass of the third body $M_3 = 0.54 \pm
0.06~$M$_\odot$ (see Table 2) and also the distance to the system V2388~Oph.

\begin{table}
 \centering
 \begin{minipage}{100mm}
  \caption{Final parameters of combined solution for V2388~Oph.}  \label{ParametersV2388Oph}
  \begin{tabular}{@{}c c c@{}}
\hline
                     &   \multicolumn{2}{c}{V2388 Oph}     \\
 Parameter           &   Present study   & Docobo \& Andrade (2013) \\
 \hline
 $JD_0 $             & 2452500.3842 (4)  &  --             \\
 $P$ [d]             & 0.80229787 (69)   &  --             \\
 $p$ [day]           & 3277.9 (22.5)     & 3290.1 (5.5)    \\
 $p$ [yr]            & 8.975 (60)        & 9.008 (15)      \\
 $A$  [day]          & 0.00102 (6)       &  --             \\
 $T$                 & 2454197.2 (27.0)  & 2454210.0 (18.3)\\
 $\omega$ [deg]      & 208.4 (3.3)       & 238.3 (10.0)     \\
 $e$                 & 0.329 (4)         & 0.327 (2)       \\
 $i$ [deg]           & 171.6 (5.2)       & 160.9 (5.0)     \\
 $\Omega$ [deg]      & 323.2 (2.8)       & 353.6 (10.0)    \\
 $a$ [mas]           & 83.0 (3.0)        & 85.3 (2.0)      \\ \hline
 $f(m_3)$ [M$_\odot$]& 7.8 (1.4) $\cdot 10^{-5}$ & --      \\
 $M_{3}$ [M$_\odot$] & 0.54 (0.06)       &  --             \\
 $a_{12}$ [AU]       & 1.26 (0.09)       &  --             \\
 $a_{3}$ [AU]        & 4.60 (1.02)       &  --             \\
 $d$ [pc]            & 70.6 (8.9)        &  --             \\
 \hline
\end{tabular}
\end{minipage}
\end{table}

At this point it is advisable to comment also the masses of the visual binary presented by Docobo
\& Andrade (2013). They gave the values $1.7$~M$_\odot$ for the primary, while $1.3$~M$_\odot$ for
the secondary of the visual double, and also a note about the magnitude difference of about 0.2~mag
only. But this is also rather doubtful value, because stars on the main sequence with such masses
should have much larger magnitude difference (see e.g. Harmanec 1988). Also the ``Delta-m
catalogue" \footnote{http://ad.usno.navy.mil/wds/dmtext.html} lists the magnitude difference about
$\Delta m = 1.5$~mag, which implies much different components of the visual pair. Our resulting
values provide a better fit to the $\Delta m$ value. The amplitude of the eclipsing pair variation
is not so large to shift the $\Delta m$ value so low. Therefore, we can speculate about the origin
of such a large magnitude change during the last 40 years suspecting that the photometric variation
comes from the third component. Such an explanation can explain why in 1960's the magnitude
difference was about 0.3~mag (van den Bos 1963), while about 40 years later this difference was
about 1 magnitude larger (Horch et al. 2010).

From our analysis the distance to the system resulted in about $70.6 \pm 8.9$~pc. The original
Hipparcos value was $67.9 \pm 4.0$~pc (Perryman et al. 1997), later recomputed to $83.3 \pm 6.1$~pc
(van Leeuwen 2007). On the other hand, Yakut et al. (2004) presented the distance $68 \pm 4$~pc,
and Docobo \& Andrade (2013) gave the two values $74.3 \pm 2.2$~pc, and $72.3 \pm 2.1$~pc,
respectively. The scatter of these values is still rather large, but our result has also high
uncertainty. This is due to the poor coverage of the LITE fit and its still not very well-defined
amplitude.

\subsection{V1031 Ori}

The system V1031 Ori (= HD 38735 = HR 2001 = HIP 27341 = MCA 22) is an Algol-type eclipsing binary,
discovered as a variable by Strohmeier \& Knigge (1961). Later, Olsen (1977) classified the star as
an eclipsing binary and gave the correct orbital period of about 3.4~day. The most detailed
analysis is that by Andersen et al. (1990), who observed the whole light curve in $uvby$ filters,
and also obtained 26 spectrograms of this multiple system. The analysis revealed that it is a
detached binary, with rather deep minima of about 0.4~mag in $V$ filter, both components are of A
spectral type, and the distance to the system is about $215 \pm 25$~pc. The more recent parallax
from the Hipparcos gave the distance $205 \pm 36$~pc (van Leeuwen 2007). Most recently, the
original data by Andersen et al. (1990) were recalculated by Torres et al. (2010).

Moreover, V1031~Ori was also discovered as a visual binary (McAlister et~al. 1983), and more than
twenty observations were carried out since its discovery. The magnitude difference between the two
visual components is of about 1.5~mag. Recent interferometric observations obtained during the last
decade revealed a rapid movement on the visual orbit, indicating rather short orbital period. The
first rough estimation of the visual orbit is a short discussion about the speckle data by Andersen
et al. (1990), who proposed a period more than 3 thousand years. A recent attempt at the orbital
solution from the available interferometric data was that by Zasche et al. (2009), who proposed a
visual orbit period of about 93~yr and a circular orbit, but with very poor data coverage. Since
then a few new observations were carried out, which can help us to better constrain the orbital
properties of the double.

A similar approach as introduced above was applied to this system; however, the task was a bit
simplified, because the distance was not computed. The distance can only be derived when both
methods (visual orbit as well as the LITE fit) have well-defined amplitudes (i.e. the amplitude of
LITE as well as the semimajor axis of the visual orbit), but this is not true in this case. The
periastron passage is only covered very poorly in both methods (see below), and our fit is still
only preliminary.

\begin{table}
 \centering
 \begin{minipage}{100mm}
  \caption{Final parameters of combined solution for V1031~Ori.}  \label{ParametersV1031Ori}
  \begin{tabular}{@{}c c c@{}}
\hline
                     &   \multicolumn{2}{c}{V1031 Ori}     \\
 Parameter           &   Present study   & Zasche et al. (2009) \\
 \hline
 $JD_0 $             & 2452500.3044 (3)  &  --             \\
 $P$ [d]             & 3.4055587 (26)    &  --             \\
 $p$ [day]           & 11432 (1252)      &  33843 (2414)   \\
 $p$ [yr]            & 31.3 (3.4)        &  92.66 (6.61)   \\
 $A$  [day]          & 0.0144 (69)       &  --             \\
 $T$                 & 2453025 (948)     &  2430580 (2849) \\
 $\omega$ [deg]      & 160.8 (27.0)      &  180.3 (28.5)   \\
 $e$                 & 0.921 (19)        &  0.001 (0.001)  \\
 $i$ [deg]           & 46.9 (13.0)       &  76.3 (8.2)     \\
 $\Omega$ [deg]      & 305.2 (11.3)      &  291.9 (9.4)    \\
 $a$ [mas]           & 94.4 (6.4)        &  176 (7)        \\ \hline
 $f(m_3)$ [M$_\odot$]& 0.13 (0.03)       &  --             \\
 $M_{3}$ [M$_\odot$] & 2.65 (0.69)       &  --             \\
 $a_{12}$ [AU]       & 6.93 (1.67)       &  --             \\
 $a_{3}$ [AU]        & 12.4 (3.0)        &  --             \\
 \hline
\end{tabular}
\end{minipage}
\end{table}

\begin{figure}
  \centering
  \includegraphics[width=\textwidth]{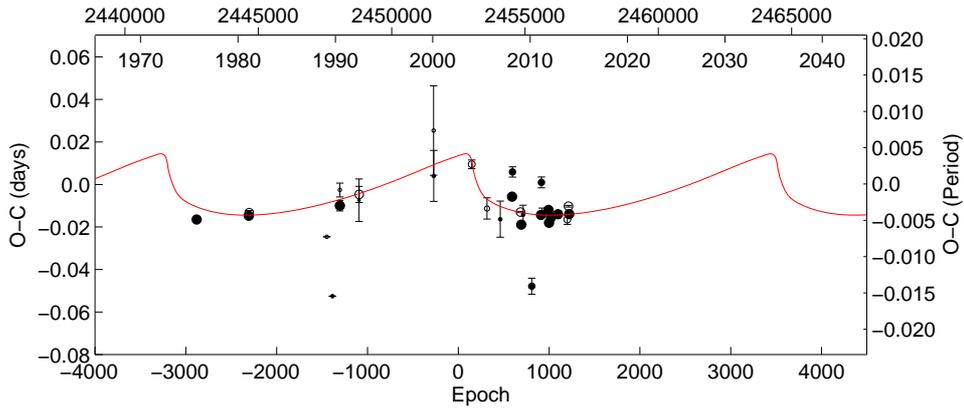}
  \caption{$O-C$ diagram of V1031 Ori.  See Fig. 1 for description of the symbols.}
  \label{FigV1031OriOC}
\end{figure}

\begin{figure}
  \centering
  \includegraphics[width=100mm]{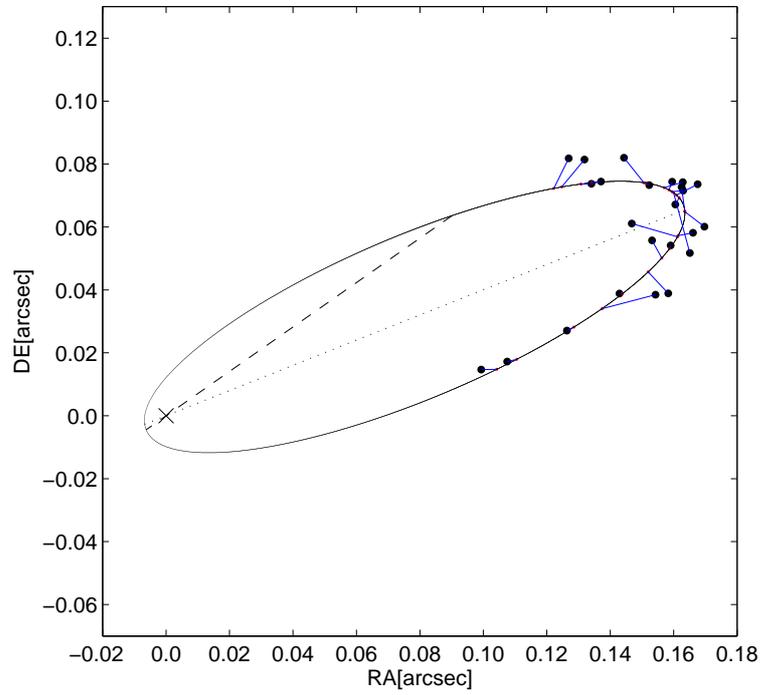}
  \caption{Visual orbit of V1031 Ori.}
  \label{FigV1031OriOrbit}
\end{figure}

\begin{figure}
  \centering
  \includegraphics[width=85mm]{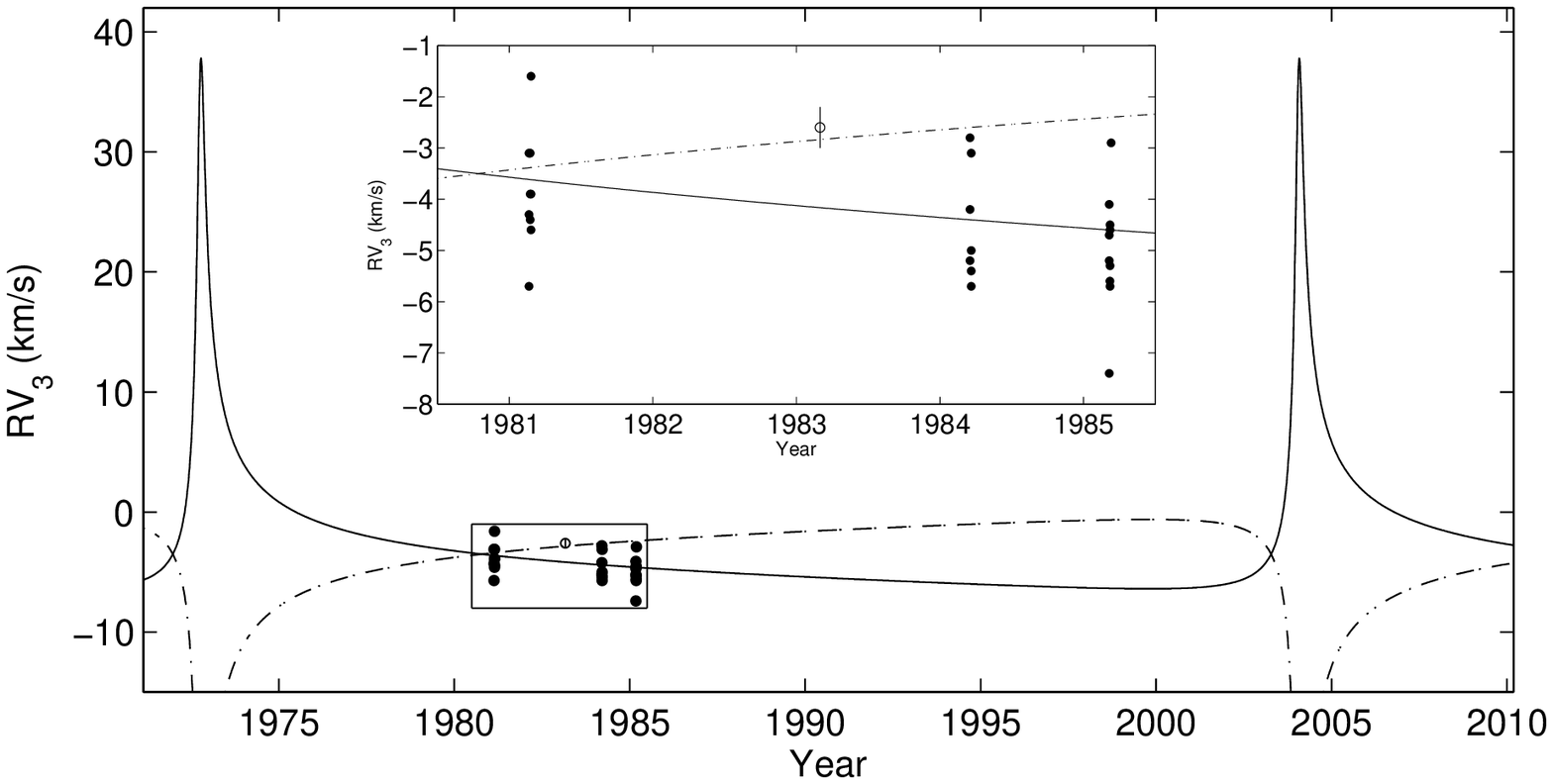}
  \caption{Radial velocities of V1031 Ori as measured by Andersen et al. (1990). The third component's velocities
  are plotted as dots, the radial velocity of barycenter of the eclipsing pair as open circle. The small rectangle is zoomed for a better clarity.}
  \label{FigV1031OriRV3}
\end{figure}

Our resulting fit is presented in Figs. 6 and 7. The parameters are given in Table 3. As one can
see, the visual orbit is very different from the one proposed in Zasche et al. (2009). This is due
to the fact that in the previous work only a small arc of the orbit was covered with observations,
and these observations were obtained away from the periastron. New observations were secured closer
to the periastron passage, hence a new orbit with better defined parameters was derived. We
observed several new minima during the last five years and also collected some already published
observations (see the Appendix tables), which show only mild additional variation. This is due to
the fact that these were obtained away from the periastron passage, where the apparent period
change of the eclipsing binary is only very slow, see Fig. 6. On the other hand, the orbit as
derived from the recent interferometric observations is defined quite well, because the binary just
completed one revolution since its discovery. Quite interesting is a value of high eccentricity of
the long orbit. Our final result about the mass of the third component of about 2.65~$M_\odot$ is
in good agreement with the previous finding by Andersen et al. (1990), who proposed a mass
2.2~$M_\odot$.

We also presented Fig. 8, where the predicted radial velocity variation is plotted together with
the observations of the third-body lines and the systemic velocity (i.e. radial velocity of the
barycenter) of the eclipsing pair as published by Andersen et al. (1990). As one can see, our fit
computed from the parameters listed in Table 3 is able to describe the observed velocities quite
well, however its conclusiveness is still poor due to only small time interval of the observations.
Hence, some new observations close to the upcoming periastron passage in 2035 would be very useful.


\section{Discussion and Conclusions}

We performed the combined analyses of the visual orbit and the apparent period changes of three
eclipsing binaries V819~Her, V2388~Oph, and V1031~Ori. These systems belong to a still rather
limited group of stars, where the eclipsing binary is a part of multiple stellar system, and the
long orbit was observed via interferometry. The long-term collecting of the spectroscopic data for
several years or decades is a bit complicated nowadays (due to time allocation policy on larger
telescopes). Hence, this method which does not need any radial velocities of the long orbit could
be a way how to work around this problem.

On the other hand, there are still only a few systems where this approach has been applied. We can
divide the group of eclipsing subsystems in the visual doubles (where the visual orbit was derived)
into three groups (see e.g. Zasche et al. 2009):

\begin{itemize}
 \item Systems, where both the visual orbit and the LITE was computed simultaneously: QS~Aql,
i~Boo, VW~Cep, KR~Com, V772~Her, V819~Her, V2388~Oph, V1031~Ori, $\zeta$~Phe, V505~Sgr, DN~UMa, and
HT~Vir.
 \item Those for which the visual orbit is known (as well as the solution of the light and
radial velocity curves), but the LITE was not detected yet: ET~Boo, V831~Cen, V2083~Cyg, MR~Del,
LO~Hya, DI~Lyn, GT~Mus, $\delta$~Ori, $\eta$~Ori, $\beta$~Per, V592~Per, V1647~Sgr, V906~Sco,
BB~Scl, $\xi$~Tau, and $\delta$~Vel.
 \item And finally stars, for which the orbit is known, but no light curve analysis (nor
the LITE analysis) was performed: V559~Cas, V773~Cas, V871~Cen, V949~Cen, BR~Ind, V635~Mon, CN~Hyi,
V410~Pup, XY~Pyx, and $\lambda$~Sco.
\end{itemize}

A future detailed analysis of all these systems would be of interest, especially for shifting the
systems from the lower two groups into the first one. The combined analysis as presented in this
paper would be very useful in this way. When having the complete set of orbital parameters for both
orbits in a particular system, one can obtain the ratio of the periods, mutual inclination of the
orbits, their eccentricities, mass ratios, etc. All of these parameters can help to better
understand the formation processes in the multiple stellar systems (see e.g. Halbwachs et al. 2003,
or Tokovinin 2008).

\Acknow{This research has made use of the Washington Double Star Catalog maintained at the U.S.
Naval Observatory. We would like to thank also Mr. Hans Zirm for a fruitful discussion about the
system V1031~Ori. An anonymous referee is also acknowledged for his/her helpful and critical
suggestions, which greatly improved the paper. This investigation was supported by the Research
Programme MSM0021620860 of the Czech Ministry of Education, by the Czech Science Foundation grant
no. P209/10/0715, and by the grant UNCE 12 of the Charles University in Prague. We also do thank
the ASAS and ``Pi of the sky" teams for making all of the observations easily public available.
This research has made use of the SIMBAD and VIZIER databases, operated at CDS, Strasbourg, France
and of NASA's Astrophysics Data System Bibliographic Services.}


\appendix

\begin{appendix}


\begin{table*}
 \tiny
 \caption{List of the minima timings used for the analysis} \label{Minima}
 \centering
\begin{tabular}{llllcl}
\hline\hline\noalign{\smallskip}
 Star       &    HJD  &  Error  & Type  & Filter & Source       \\
 designation&   - 2400000   &  [day]  &       &        & Observatory  \\
\noalign{\smallskip}\hline \noalign{\smallskip}
  V819~Her & 45169.8078   & 0.0050  & Sec   &   V    & Wasson et al. (1994)  \\  
  V819~Her & 45169.8183   & 0.0040  & Prim  &   B    & Wasson et al. (1994)  \\  
  V819~Her & 45471.9262   & 0.0018  & Prim  &   U    & Wasson et al. (1994)  \\  
  V819~Her & 45471.9280   & 0.0014  & Prim  &   B    & Wasson et al. (1994)  \\  
  V819~Her & 45471.9282   & 0.0040  & Prim  &   V    & Wasson et al. (1994)  \\  
  V819~Her & 45519.8735   & 0.0033  & Sec   &   B    & Wasson et al. (1994)  \\
\hline
\end{tabular}
 \begin{list}{}{}
 \item[] Columns: [1] - The standard star designation, [2] - Heliocentric Julian date of time of
 mid-eclipse, [3] - Standard error of the time of minimum, [4] - Primary/Secondary minimum type,
 [5] - Photometric filter used, [6] - Reference of the particular observation.
 \end{list}
\end{table*}

\begin{table*}
 \tiny
 \caption{List of the minima timings used for the analysis} 
 \centering
\begin{tabular}{llllcl}
\hline\hline\noalign{\smallskip}
 Star       &    HJD  &  Error  & Type  & Filter & Source       \\
 designation&   - 2400000   &  [day]  &       &        & Observatory  \\
\noalign{\smallskip}\hline \noalign{\smallskip}
  V819~Her & 48486.3889   & 0.0040  & Prim  &   V    & Wasson et al. (1994)  \\  
  V819~Her & 48502.0100   & 0.001   & Prim  &   Hp   & Perryman et al. (1997)                  \\  
  V819~Her & 48505.3475   & 0.0035  & Sec   &   V    & Wasson et al. (1994)  \\  
  V819~Her & 48780.710    & 0.002   & Prim  &   V    & Wasson et al. (1994)  \\  
  V819~Her & 48780.711    & 0.001   & Prim  &   B    & Wasson et al. (1994)  \\  
  V819~Her & 48780.7136   & 0.0003  & Prim  &   C    & Wasson et al. (1994)  \\
\hline
\end{tabular}
\end{table*}

\begin{table*}
 \tiny
 \caption{List of the minima timings used for the analysis} 
 \centering
\begin{tabular}{llllcl}
\hline\hline\noalign{\smallskip}
 Star       &    HJD  &  Error  & Type  & Filter & Source       \\
 designation&   - 2400000   &  [day]  &       &        & Observatory  \\
\noalign{\smallskip}\hline \noalign{\smallskip}
  V2388~Oph& 49890.5045   &         & Prim  &  uvby  & Rodr{\'{\i}}guez et al. (1998)   \\  
  V2388~Oph& 51569.7076   & 0.0006  & Prim  &   BVR  & Yakut et al. (2004)   \\  
  V2388~Oph& 51700.4852   & 0.0009  & Prim  &   BVR  & Yakut et al. (2004)   \\  
  V2388~Oph& 51749.430    & 0.001   & Prim  &   BVR  & Yakut et al. (2004)   \\  
  V2388~Oph& 51780.320    & 0.002   & Sec   &   BVR  & Yakut et al. (2004)   \\  
  V2388~Oph& 52816.88927  & 0.00373 & Sec   &    V   & ASAS - This paper           \\
\hline
\end{tabular}
\end{table*}

\begin{table*}
 \tiny
 \caption{List of the minima timings used for the analysis} 
 \centering
\begin{tabular}{llllcl}
\hline\hline\noalign{\smallskip}
 Star       &    HJD  &  Error  & Type  & Filter & Source       \\
 designation&   - 2400000   &  [day]  &       &        & Observatory  \\
\noalign{\smallskip}\hline \noalign{\smallskip}
  V2388~Oph& 55830.32345  & 0.00045 & Sec   &   R    & This paper                  \\  
  V2388~Oph& 55993.58942  & 0.00031 & Prim  &   V    & This paper                  \\  
  V2388~Oph& 56040.52504  & 0.00080 & Sec   &   I    & This paper                  \\  
  V2388~Oph& 56073.41700  & 0.00053 & Sec   &   I    & This paper                  \\  
  V2388~Oph& 56095.48013  & 0.00024 & Prim  &   R    & This paper                  \\  
  V2388~Oph& 56097.48636  & 0.00058 & Sec   &   I    & This paper                  \\
\hline
\end{tabular}
\end{table*}


\begin{table*}[h]
 \tiny
 \caption{Positional measurements used for the analysis.}
  \centering
 \begin{tabular}{l l l r l}
 \hline\hline
 System    & Year  & $\rho$ [arcs] & $\phi$ [deg] & Reference \\
 \hline
 V819 Her  &   1980.4766   &   0.036   &  278.1  & McAlister et al. (1983) \\  
 V819 Her  &   1980.4820   &   0.047   &  265.7  & McAlister et al. (1983) \\  
 V819 Her  &   1981.4568   &   0.065   &  134.1  & McAlister et al. (1984) \\  
 V819 Her  &   1981.4678   &   0.061   &  134.2  & McAlister et al. (1984) \\  
 V819 Her  &   1981.4706   &   0.063   &  135.2  & McAlister et al. (1984) \\  
 V819 Her  &   1981.4732   &   0.062   &  135.6  & McAlister et al. (1984) \\  
 V819 Her  &   1982.2924   &   0.107   &  154.7  & Fu et al. (1997) \\
\hline
 \end{tabular}
 \end{table*}

\begin{table*}[h]
 \tiny
 \caption{Positional measurements used for the analysis.}
  \centering
 \begin{tabular}{l l l r l}
 \hline\hline
 System    & Year  & $\rho$ [arcs] & $\phi$ [deg] & Reference \\
 \hline
 V819 Her  &   2004.52234  &   0.10308 &  153.38 & Muterspaugh et~al. (2008)     \\  
 V819 Her  &   2004.52502  &   0.10322 &  153.35 & Muterspaugh et~al. (2008)     \\  
 V819 Her  &   2004.5265   &   0.10308 &  153.38 & Muterspaugh et al. (2010) \\  
 V819 Her  &   2004.52795  &   0.10194 &  154.41 & Muterspaugh et~al. (2008)     \\  
 V819 Her  &   2004.5292   &   0.10322 &  153.35 & Muterspaugh et al. (2010) \\  
 V819 Her  &   2004.5320   &   0.10194 &  154.41 & Muterspaugh et al. (2010) \\  
 V819 Her  &   2004.55249  &   0.10349 &  153.34 & Muterspaugh et~al. (2008)     \\
\hline
 \end{tabular}
 \end{table*}

\begin{table*}[h]
 \tiny
 \caption{Positional measurements used for the analysis.}
  \centering
 \begin{tabular}{l l l r l}
 \hline\hline
 System    & Year  & $\rho$ [arcs] & $\phi$ [deg] & Reference \\
 \hline
 V2388 Oph &   1986.4050   &   0.097   &  258.8  & Balega et al. (1989) \\ 
 V2388 Oph &   1986.4100   &   0.099   &  257.9  & McAlister et al. (1989) \\ 
 V2388 Oph &   1986.6563   &   0.095   &  249.8  & Balega et al. (1989) \\ 
 V2388 Oph &   1987.2700   &   0.082   &  231.8  & McAlister et al. (1989) \\ 
 V2388 Oph &   1987.3724   &   0.078   &  226.1  & Balega et al. (1989) \\ 
 V2388 Oph &   1988.2611   &   0.063   &  190.5  & McAlister et al. (1989) \\ 
 V2388 Oph &   1988.6655   &   0.056   &  157.1  & McAlister et al. (1990) \\
\hline
 \end{tabular}
 \end{table*}

\end{appendix}

\end{document}